\documentclass[twocolumn]{autart}

\usepackage{graphicx}
\usepackage{amsmath}
\usepackage{amssymb}
\usepackage{booktabs} 
\usepackage[authoryear,longnamesfirst]{natbib}
\usepackage[bookmarks=false]{hyperref}
\usepackage{booktabs}
\usepackage{csvsimple}
\usepackage{siunitx}
\usepackage{ifthen} 
\usepackage{natbib}

\newcommand{\MaybeInfinity}[1]{%
  \ifthenelse{\equal{#1}{Inf} \OR \equal{#1}{NaN}}{$\infty$}{#1}%
}

\newtheorem{remark}{Remark}
\newtheorem{theorem}{Theorem}

\begin{document}

\begin{frontmatter}

\title{A Comparison of Set-Based Observers for Nonlinear Systems}


\author[TUM]{Nico Holzinger*}\ead{nico.holzinger@tum.de},
\author[TUM]{Matthias Althoff}\ead{althoff@tum.de}

\address[TUM]{Department of Computer Engineering, Technical University of Munich, Garching, Germany}


\begin{abstract}
Set-based state estimation computes sets of states consistent with a system model given bounded sets of disturbances and noise. Bounding the set of states is crucial for safety-critical applications so that one can ensure that all specifications are met. While numerous approaches have been proposed for nonlinear discrete-time systems, a unified evaluation under comparable conditions is lacking. This paper reviews and implements a representative selection of set-based observers within the CORA framework. To provide an objective comparison, the methods are evaluated on common benchmarks, and we examine computational effort, scalability, and the conservatism of the resulting state bounds. This study highlights characteristic trade-offs between observer categories and set representations, as well as practical considerations arising in their implementation. All implementations are made publicly available to support reproducibility and future development. This paper thereby offers the first broad, tool-supported comparison of guaranteed state estimators for nonlinear discrete-time systems.
\end{abstract}

\end{frontmatter}

\section{Introduction}
Reliable knowledge of the system state is crucial for safety-critical cyber-physical systems. Yet in practice, only a subset of states can be measured directly due to sensor limitations, cost constraints, or spatial restrictions. In addition, the available measurements are typically affected by noise. While for non-critical applications, state estimation using probabilistic estimation techniques like Kalman filters is sufficient, for safety-critical applications, guaranteed bounds of the state are required. These safety-critical applications include surgical robots (\cite{haidegger_autonomy_2019}), autonomous driving (\cite{althoff_online_2014}), or human-robot interaction in rehabilitation (\cite{yu_humanrobot_2015}) to name a few. 

To obtain guaranteed bounds for the state, set-based state estimation is used. The basic idea is to obtain a set of states that is consistent with the system model and the available measurements considering bounded noise and disturbances, thereby guaranteeing the inclusion of the real state in the estimated set. These guaranteed sets can be used for robust control (\cite{schurmann_reachset_2018}), to rigorously predict future behaviors of a system (\cite{althoff_set_2021}) or for robust fault detection (\cite{chabane_fault_2015,mu_set-based_2024}).

One deciding difference between approaches is the chosen set representation. Methods relying on intervals have been investigated in \cite{kieffer_guaranteed_1998,jaulin_nonlinear_2002,yang_accurate_2018} while ellipsoidal methods were developed in \cite{gollamudi_set-membership_1998,scholte_nonlinear_2003,liu_ellipsoidal_2016,wang_dual_2022}. More recently, zonotope-based approaches became popular (\cite{combastel_state_2005,alamo_guaranteed_2005,le_zonotopic_2013}), which have been extended to zonotope bundles (\cite{althoff_zonotope_2011,khajenejad_guaranteed_2021}) or constrained zonotopes (\cite{scott_constrained_2016,rego_set-based_2018}).

 While the pioneering works of set-based estimation focused on linear time-invariant systems (\cite{schweppe_recursive_1968,bertsekas_recursive_1971}), recent research has increasingly addressed methods for nonlinear systems (\cite{alamo_guaranteed_2005,combastel_state_2005,scott_constrained_2016,yang_efficient_2018}). A detailed review of these nonlinear approaches is provided in Sections~\ref{intersection_based}, \ref{propagation_based}, and \ref{interval_observers}. Obtaining tight enclosures for nonlinear dynamics remains challenging, and despite several promising proposals, the field still lacks unified benchmarks. A growing interest in guaranteed state estimation has been observed over the past years, and some works have compared the performance of their observers with previously developed ones. \cite{rego_set-based_2025} and \cite{yang_set-membership_2025} both benchmark their proposed methods against two existing observers on up to four-dimensional systems to demonstrate the advantages of their approach. In \cite{mu_comparison_2022}, the authors compare their new method together with three set-based approaches for fault detection of nonlinear systems to more conventional fault detection methods like the EKF, concluding that set-based methods offer clear potential for fault detection. Furthermore, a new toolbox for set-based estimation in MATLAB called ZETA has recently been introduced  (\cite{rego_zeta_2025}), which offers observers for nonlinear systems based on zonotopes and constrained zonotopes. Nevertheless, to the best of our knowledge, no comprehensive comparison across approaches using different set representations has been conducted, and no openly available benchmark suite exists. This work closes this gap by implementing and comparing a broad variety of set-based estimation approaches for nonlinear discrete-time systems.

In a comparative work for linear set-based observers by \cite{althoff_comparison_2021}, the estimation techniques are split in three categories. First, strip-based methods use strips to represent the measurement information and intersect them with the predicted set to bound the set of possible states. Second, propagation-based methods propagate sets through a Luenberger observer to update the predicted set to avoid intersecting sets. Third, interval observers separately propagate lower and upper bounds for each state component. For the comparison of observers for nonlinear systems, we will use the same categorization. However, not all intersection-based methods rely on measurement strips, especially considering possible future extensions to nonlinear measurement equations. Furthermore, while \cite{efimov_interval_2013} proposed a true interval observer for a specific class of continuous-time nonlinear systems, this construction does not readily extend to general discrete-time dynamics. To the best of the authors' knowledge, no pure interval observers exist for general nonlinear discrete-time systems. However, there are methods that propagate upper and lower bounds separately in the prediction step and then additionally use intersection in the correction step. While interval observers for linear systems do not require intersection, we still categorize observers for nonlinear systems as interval observers if the reachable set is obtained from separately computing a lower and an upper bound.

 Our contributions are the comparison of existing, peer-reviewed methods as well as the implementation of the investigated approaches in the open-source MATLAB toolbox CORA (\cite{althoff_introduction_2015}) for further comparison of new approaches and easy reproducibility. The remainder of this paper is organized as follows. Section~\ref{preliminaries} summarizes the theoretical foundations of set-based computations and introduces the investigated observer categories. Section~\ref{comparison} describes the comparison methodology and the used benchmarks. In Section~\ref{results}, we present the obtained results, and Section~\ref{conclusion} provides concluding remarks.

\section{Preliminaries}
\label{preliminaries}
This section introduces the required mathematical background, including set representations and operations commonly used in guaranteed state estimation. Additionally, we provide an overview of the compared methods and their categorization.

\subsection{Set Representations}
Before the considered types of sets are introduced, we define some basic set operations that are used throughout this paper. Considering sets $\mathcal{X},\mathcal{W} \subset \mathbb{R}^n$, $\mathcal{Y} \subset \mathbb{R}^m$ and a real matrix $C \subset \mathbb{R}^{n \times m}$, the basic set operations Minkowski sum, linear map and intersection are defined as \cite[Def. 2]{scott_constrained_2016}
\begin{align}
\mathcal{X} \oplus \mathcal{W} &:= \{x + w \mid x \in \mathcal{X}, w \in \mathcal{W}\}, \\
 C \mathcal{X} &:= \{C x \mid x \in \mathcal{X}\}, \\
 \mathcal{X} \cap_C \mathcal{Y} &:= \{x \in \mathcal{X} \mid C x \in \mathcal{Y}\},
\end{align}    
respectively. The most basic set representation used in this paper is an interval, which is defined as 
$[\underline{x},\overline{x}] = \{ x \in \mathbb{R} \mid \underline{x} \le x \le \overline{x} \}$. 
An interval vector represents an axis-aligned box as 
$[[\underline{x}_1,\overline{x}_1],[\underline{x}_2,\overline{x}_2], \ldots, [\underline{x}_n,\overline{x}_n]]^T$. 
A unit interval is defined as $\mathcal{B} = [-1, 1]$ and the unit box as $\mathcal{B}^n \subset \mathbb{R}^n$. 
We denote by $\mathbb{IR}^n$ the space of $n$-dimensional real interval vectors.

Another simple set representation is the ellipsoid, which is defined as \cite[Def. 1]{scholte_nonlinear_2003}
\begin{equation}
    \Omega = \left\{ x \,\middle|\, (x-a)^{\!T} P^{-1} (x-a) \le 1 \right\},
\end{equation}
where $a \in \mathbb{R}^n$ is the center and $P \in \mathbb{R}^{n \times n}$ the shape matrix of the ellipsoid. While ellipsoids can be represented efficiently even for higher dimensions, they are not closed under Minkowski sum or intersection and are limited in representing complex shapes. Another important set representation are polyhedra, which, in H-representation, are defined as the intersection of a finite number of half-spaces \cite[Def. 3.26]{chabane_fault_2015}:
\begin{equation}
    \mathcal{P} = \{ x \in \mathbb{R}^n \mid H x \le h \},
\end{equation}
where the row vectors of $H\in \mathbb{R}^{m \times n}$ are the normal vectors to the halfspaces and $h\in \mathbb{R}^m$ determines the distance of the halfspaces to the origin. A bounded polyhedron is called a polytope. While polyhedra can represent more complex shapes than intervals and are closed under Minkowski sums and intersections, their manipulation becomes computationally expensive in high dimensions; for instance, the Minkowski sum of two polytopes in H-representation has a complexity of $\mathcal{O}(2^n)$ \cite[Tab. 1]{althoff_comparison_2021}.
 Zonotopes are a special class of polytopes that can be represented as
\begin{equation}
    \mathcal{Z} = \{ c + G \xi \mid \xi \in \mathcal{B}^r \},
\end{equation}
where $c \in \mathbb{R}^n$ is the center and $G \in \mathbb{R}^{n \times r}$ the generator matrix. The order of a zonotope is defined as $r/n$ (\cite{scott_constrained_2016}). Zonotopes are closed under Minkowski sum and linear maps with a complexity of $\mathcal{O}(n)$ and $\mathcal{O}(n^3)$ respectively \cite[Tab. 1]{althoff_comparison_2021}, making them suitable for high-dimensional systems. However, they are limited to centrally symmetric shapes and are not closed under intersection. One way to use the concept of zonotopes for more general convex sets is to use the intersection of multiple zonotopes to obtain zonotope bundles (\cite{althoff_zonotope_2011}). The last set representation considered in this paper are constrained zonotopes, which extend zonotopes by adding linear equality constraints to reduce conservatism. A constrained zonotope is defined as \cite[Def. 3]{scott_constrained_2016}
\begin{equation}
    \mathcal{CZ} = \{ c + G \xi \mid \xi \in \mathcal{B}^r, A \xi = b \},
\end{equation}
where $A \in \mathbb{R}^{q \times r}$ and $b \in \mathbb{R}^q$ define the linear constraints. Constrained zonotopes can represent general polytopes and are closed under linear maps, Minkowski sums and intersection which can all be computed efficiently using matrix operations as shown in \cite{scott_constrained_2016}. The practical bottleneck is the growth of the generator and constraint matrices under repeated operations, which necessitates 
order and constraint reduction techniques to keep computations tractable. The original reduction techniques in \cite{scott_constrained_2016} have recently been improved by \cite{rego_novel_2025} to reduce overall complexity significantly.

\subsection{Range Bounding}
For a nonlinear function $f : \mathbb{R}^n \to \mathbb{R}^m$, an inclusion function $F : \mathbb{IR}^n \to \mathbb{IR}^m$ guarantees
\[
f(\mathcal{Z}) = \{\, f(z) \mid z \in \mathcal{Z} \,\} \subseteq F(\mathcal{Z})
\quad \text{for all } \mathcal{Z} \in \mathbb{IR}^n.
\]
Such inclusion functions can be computed using interval arithmetic and provide 
guaranteed enclosures of nonlinear mappings (\cite{alamo_guaranteed_2003}). Alternative range bounding methods are reviewed in \cite{althoff_range_bounding_2018}. A widely used method for range bounding is:

\begin{theorem}[Mean Value Extension]\cite[Theorem 2]{alamo_guaranteed_2005}
  \label{MV_theorem}
Consider a continuously differentiable function $f:\mathbb{R}^n\to\mathbb{R}$, an interval vector $\mathcal{X}\in\mathbb{IR}^n$, and let $c\in \mathcal{X}$.
Then
\[
f(X) \subseteq f(c) \oplus \big(\nabla f(\mathcal{X})\big)(\mathcal{X}-c),
\]
where $\nabla f(\mathcal{X})$ denotes an interval enclosure of the gradient of~$f$ over~$\mathcal{X}$.
\end{theorem}

\subsection{Problem Statement}
We consider nonlinear discrete-time systems with additive disturbance and a linear measurement equation of the form
\begin{align}
x_{k+1} &= f(x_k,u_k) + w_k, \label{nonlinearSys}\\
y_k &= Cx_k + v_k, \label{measeq}
\end{align}
where $x_k \in \mathbb{R}^n$ is the state, $u_k \in \mathbb{R}^m$ the input and $C \in \mathbb{R}^{r \times n}$ is the measurement matrix. The disturbance and measurement noise are bounded by $w_k \in \mathcal{W}$ and $v_k \in \mathcal{V}$.

For a given state $x_k$, input $u_k$, and disturbance $w_k$, 
we denote the next state of (\ref{nonlinearSys}) by $\chi(x_k, u_k, w_k)$. 
Our goal is to characterize the sets of all possible states at time step~$k$, 
starting from the initial set $\mathcal{S}_0 \subset \mathbb{R}^n$ as
\begin{align}
    \mathcal{S}_k = \{&x_k \mid x_{k-1} \in \mathcal{S}_{k-1}, w_{k-1} \in \mathcal{W}, v_k \in \mathcal{V},\\
    & x_k = \chi(x_{k-1}, u_{k-1}, w_{k-1}), y_k = Cx_k + v_k\}.\nonumber
\end{align}
Our objective is to compute a tight over-approximation of $\mathcal{S}_k$. The three main categories and their subcategories of methods are introduced in the following sections. 

\begin{remark}
    While the focus of this paper is on systems with additive noise and linear measurements, many of the discussed methods can be extended to more general settings, including multiplicative noise and nonlinear measurement equations. However, for a clearer comparison, and since many practical examples are covered by these systems, we restrict ourselves to this common setting.
\end{remark}

\subsection{Intersection-based Methods}
\label{intersection_based}
Intersection-based methods use prediction and correction. In the prediction step, the reachable set is computed based on the system dynamics and disturbances $\mathcal{W}$. Next, in the correction step, the reachable set is intersected with the set of states consistent with the measurement.

 While the underlying concept is similar across intersection-based methods, various intersection techniques have been developed to improve tightness and efficiency. For the linear measurement equation in \eqref{measeq}, each measurement $y_{k,j}$ together with the bounded noise $v_k \in \mathcal{V}$
defines a measurement strip of the form
\[
\hat S_j = \{\, x \in \mathbb{R}^n \mid C_j x - y_{k,j} \in [\underline{v}_j,\, \overline{v}_j] \,\},
\]
where $C_j$ is the $j$th row of $C$ and $[\underline{v}_j,\,\overline{v}_j]$
is the noise bound for the $j$th measurement.
These strips can then be intersected with the predicted reachable set,
and efficient over-approximative intersection operators for zonotopes are
discussed in Sec.~3.1 of~\cite{althoff_comparison_2021}. For constrained zonotopes, the intersection can be performed directly as demonstrated in \cite{scott_constrained_2016}. As the correction step has already been detailed in other works, we focus more on computing the reachable set, especially since it cannot be computed exactly for most nonlinear systems.

\paragraph*{Mean value extension.}
A quite common technique applied for intersection-based observers is based on Theorem \ref{MV_theorem}. Several proposed estimators use this principle, such as FRad-A and VolMin-A (\cite{alamo_guaranteed_2005}), which use zonotopes and minimize the Frobenius norm and volume of the estimated set respectively, 
as well as a more recent method by \cite{rego_guaranteed_2020}, 
which adapts the formulation to constrained zonotopes and is denoted as CZMV.

\paragraph*{Difference of Convex Functions.}
Another family of intersection-based methods expresses the dynamics as the difference of convex (DC) functions, $f(x_k,u_k) = g(x_k,u_k) - h(x_k,u_k)$,
where both $g$ and $h$ are convex functions in the considered domains. A broad class of nonlinear functions is expressible in this form, for example every twice continuously differentiable function on a compact set (\cite{alamo_set-membership_2008}). However, finding a good decomposition is in general non-trivial and often requires manual tuning.
This decomposition allows one to bound $f$ tightly by using supporting tangents of the convex functions $g$ and $h$: a tangent provides a global lower bound for each of $g$ and $h$, and these bounds can be used to obtain tight lower and upper bounds on $f = g - h$.
The DC-based observer was first introduced for zonotopes (ZDC)~(\cite{alamo_set-membership_2008}) 
and later extended to constrained zonotopes (CZDC)~(\cite{de_paula_set-based_2024}).

\paragraph*{Conservative linearization methods.}
Several set-based observers for nonlinear systems rely on conservative linearization, that is, on linearizing the nonlinear dynamics and over-approximating the higher-order terms using the Hessian of the dynamics. While this idea is presented for continuous-time systems in \cite{althoff_reachability_2008}, the idea applies equally to discrete-time settings. An early example is the nonlinear set-membership filter (ESO-E) of \cite{scholte_nonlinear_2003}, where the reachable set is represented by ellipsoids and propagated via a linear approximation of the dynamics together with a bound on the Taylor series remainder. The correction step is then obtained by intersecting the predicted ellipsoid with a measurement strip and computing a minimum-volume enclosing ellipsoid of the intersection.

A similar idea underlies several methods that were originally designed for linear systems but can be applied to nonlinear systems through conservative linearization (\cite{althoff_guaranteed_2021}). In these approaches, the linearized dynamics are used to propagate the chosen set representation, while the Hessian is used to over-approximate the higher-order terms to ensure guaranteed inclusion of the true reachable set. The FRad-B method (\cite{wang_set-membership_2018}), like FRad-A, minimizes the Frobenius norm of the generator matrix, but employs a more advanced intersection technique. VolMin-B (\cite{bravo_bounded_2006}) minimizes the volume analogously to VolMin-A and likewise differs in its correction step. The CZN-A (\cite{scott_constrained_2016}) and CZN-B (\cite{alanwar_privacy-preserving_2023}) methods use constrained zonotopes: CZN-A performs an exact constrained zonotope intersection for the correction step, whereas CZN-B relies on measurement strip intersection. An overview of all considered methods is given in Table~\ref{tab:methods}.

\subsection{Propagation-based Methods}
\label{propagation_based}
In propagation-based methods, the estimated sets are calculated by a set-based evaluation of a Luenberger observer, avoiding the need for intersection. This is achieved by computing a correction matrix $G$ which, multiplied with the error between measurements and prediction, corrects the estimated set. Only one approach for nonlinear systems could be found in the literature that belongs to this category, which are introduced subsequently.

\paragraph*{Conservative linearization method.}
While there are multiple propagation-based approaches for linear systems (\cite{althoff_guaranteed_2021}), FRad-C (\cite{combastel_zonotopes_2015}) is the only one surveyed in the previous work that relies on online computation of the correction matrix, which makes conservative linearization possible for this approach. The Luenberger observer on which this method is based is given by
\begin{equation}
    x_{k+1} = Ax_k + Bu_k + w_k + G(y_k - Cx_k - v_k).
    \label{eq:Luenberger}
\end{equation}

It is shown in Prop. 1 of \cite{combastel_zonotopes_2015} that a set-valued extension of this observer yields an outer approximation of the true state when point-valued variables are replaced by bounded sets. The computation of $G$ is based on an adapted Kalman filter formulation minimizing the Frobenius norm of the resulting set. This is done online at each time step, after the nonlinear system has been linearized conservatively.

\subsection{Interval Observers}
\label{interval_observers}
As mentioned in the introduction, the following observers do not precisely satisfy the original definition of interval observers (\cite{GOUZE200045}) since they use intersections. Still they are listed here since they maintain lower and upper bounds for each state component separately.

\paragraph*{Differential inequalities.}
These estimators use differential inequalities (DI) to propagate guaranteed interval enclosures, exploiting the fact that, for continuous-time systems, a trajectory cannot leave a set without first touching its boundary. While this property does not generally hold for discrete-time systems, it can still be applied under certain conditions. In the original continuous-time formulation, each state component is bounded by a pair of differential inequalities
\[
\dot{x}_i^{L} \le \dot{x}_i \le \dot{x}_i^{U},
\]
whose integration yields lower and upper trajectories enclosing all possible states. 
For the discrete-time version (DTDI), the idea remains similar and it can be shown that for Euler-discretized dynamics with sufficiently small step sizes, where the bound is determined by Lipschitz constants, the guarantees carry over (\cite{yang_accurate_2018}).
Formally, the prediction equations take the form
\[
x_{k+1,i}^{L/U}
  = x_{k,i}^{L/U}
  + h\, f_i^{L/U}\!\big(I[\beta_i^{L/U}(X_k), \mathcal{G}], \mathcal{W}\big),
\]
where the face selection operators $\beta_i^{L/U}$ select the relevant interval faces,
$I[\cdot,\mathcal{G}]$ refines them using invariant or redundant constraints $\mathcal{G}$,
and $f_i^{L/U}$ are inclusion functions computed by interval arithmetic (\cite{yang_accurate_2018}). In the correction step, the upper and lower bounds are combined to form interval enclosures which are intersected with measurement strips to refine the estimates.

The DTDI approach provides tight, low-cost enclosures but requires a sufficiently small step size and is only applicable to Euler-discretized systems to ensure that these discrete inequalities remain valid. Therefore, it is not considered in the general comparison. However, the pseudo-DTDI (pDTDI) (\cite{mu_set-based_2024}) eliminates this restriction by replacing exact face selection with a one-dimensional partitioning of each coordinate direction.
Each slab is propagated separately and re-combined to form the next interval enclosure, achieving similar accuracy while making it possible to apply the observer to arbitrary discrete-time systems.

\paragraph*{Mixed-monotone decomposition.}
In this last subcategory, the propagation is performed in the generator space of constrained zonotopes or zonotope bundles. This has the advantage that their generator coefficients~$\xi$ always satisfy $\xi \in [-1,1]^{n_g}$ and by rewriting the dynamics of the system as the lifted map $\tilde{f}(\xi) = f(c + G\xi)$, the problem reduces to bounding $\tilde{f}$ over the interval domain $\xi \in [-1,1]^{n_g}$. On interval domains, recently developed mixed-monotone decomposition functions provide tight upper and lower bounds on each component of~$\tilde f$ (\cite{khajenejad_guaranteed_2021}).

A mapping $f_d : \mathcal{Z} \times \mathcal{Z} \to \mathbb{R}^m$ is called a discrete-time mixed-monotone decomposition function of $f$ according to Def. 2 in \cite{khajenejad_guaranteed_2021} if (i) $f_d(x,x) = f(x)$, (ii) it is monotone increasing in its first argument, and (iii) monotone decreasing in its second. When applied to the interval domain $[-1,1]^{n_g}$ in generator space, the corresponding lower and upper decomposition functions yield componentwise bounds on $\tilde f(\xi)$. These bounds are then used to construct a new constrained zonotope in the CZKH approach or a zonotope bundle in the ZBKH approach, which over-approximates the propagated state set. Since we restrict attention to linear measurement equations, the update step reduces to intersecting the propagated sets with the corresponding measurement strips. Next, we present how we have compared the various types of observers.

\begin{table}[t]
\centering
\caption{Overview of guaranteed state estimation methods}
\label{tab:methods}
\renewcommand{\arraystretch}{1.1}
\setlength{\tabcolsep}{5pt}
\begin{tabular}{@{}l l @{\,\hspace{10pt}} p{3.8cm}@{}}
\toprule
\textbf{Method} & \textbf{Set Rep.} & \textbf{Reference} \\ \midrule
\multicolumn{3}{l}{\textbf{Intersection-based Methods}} \\
ESO-E & Ellipsoid & \cite{scholte_nonlinear_2003}\\ 
FRad-A & Zonotope & \cite{alamo_guaranteed_2005}\\
FRad-B & Zonotope & \cite{wang_set-membership_2018}\\
VolMin-A & Zonotope & \cite{alamo_guaranteed_2005}\\
VolMin-B & Zonotope & \cite{bravo_bounded_2006}\\
ZDC & Zonotope & \cite{alamo_set-membership_2008}\\
CZDC & Constr. zonotope & \cite{de_paula_set-based_2024}\\
CZN-A & Constr. zonotope & \cite{scott_constrained_2016}\\
CZN-B & Constr. zonotope & \cite{alanwar_privacy-preserving_2023}\\
CZMV & Constr. zonotope & \cite{rego_guaranteed_2020}\\
[5pt]

\multicolumn{3}{l}{\textbf{Propagation-based Method}} \\
FRad-C & Zonotope & \cite{combastel_zonotopes_2015}\\ 
[5pt]

\multicolumn{3}{l}{\textbf{Interval Methods}} \\
pDTDI & Interval & \cite{mu_set-based_2024}\\
CZKH & Constr. zonotope & \cite{khajenejad_guaranteed_2021}\\
ZBKH & Zonotope bundle & \cite{khajenejad_guaranteed_2021}\\

\bottomrule
\end{tabular}
\end{table}

\subsection{Structural Limitations of Observer Categories}

In the following, we discuss structural limitations of the considered observer categories with respect to their ability to represent the exact consistent state set~$\mathcal{S}_k$ defined in~\ref{eq:consistent_set}. Under the assumption that the propagation through the nonlinear map~$\chi(\cdot)$ in~(\ref{nonlinearSys}) is performed optimally, we call a method \emph{optimal} if it can represent $\mathcal{S}_k$ without additional over-approximation.

Intersection-based observers explicitly construct an enclosure of $\mathcal{S}_k$ by combining the propagated set with the measurement-consistent set induced by~(\ref{measeq}). If the employed set representation admits exact or arbitrarily tight intersection operations, these methods can, in principle, represent $\mathcal{S}_k$ exactly.

Propagation-based observers, in contrast, do not perform an explicit intersection with the measurement-consistent set. Instead, the estimate is obtained via affine set-valued updates involving linear maps and Minkowski sums. Consequently, the resulting estimate is restricted to the closure of the chosen set representation under these operations. Even if the propagation through~$\chi(\cdot)$ is exact, this class of sets cannot represent general consistent state sets. For example, if the propagated set is a Euclidean ball and the measurement-consistent set is a strip, their intersection contains both curved and flat boundary segments, which cannot be generated by affine transformations and Minkowski sums alone. Hence, propagation-based observers are not optimal in this sense in general.

Interval-based observers considered in this work do not share this structural limitation, since lower and upper bounds are propagated separately and subsequently intersected with the measurement-consistent set. Therefore, interval-based approaches, particularly in combination with more expressive set representations like CZKH, can in principle achieve optimal representations of~$\mathcal{S}_k$ under exact sub-operations.

\section{Comparison}
\label{comparison}
The following subsections motivate the selection of the benchmarks with respect to nonlinearity and dimensionality and describe the evaluation criteria used to assess computational efficiency, scalability, and conservatism of the investigated approaches.

\subsection{General Setup}
All algorithms are implemented in CORA (v2026.0.1) in MATLAB (R2024b), ensuring consistent data structures and numerical routines and will be made available in a future release. The observers are implemented in a way that they can be applied to arbitrary systems of the form (\ref{nonlinearSys}). To minimize possible faults, examples from the original papers have been reproduced, and authors have been contacted for feedback on the implementations. All simulations are conducted on an AMD Ryzen 7 with 3.8 GHz and 64 GB RAM. For the optimization problems, Mosek (\cite{mosek}) was used as the solver.
For a fair comparison, identical initial, disturbance and noise sets are used in the form of intervals across all observers. In the case of the ellipsoidal observer, which cannot exactly represent intervals, the sets are converted to an enclosing ellipsoid. Simulations are conducted for 100 time steps.
Observer options such as zonotope order or constraint limits for constrained zonotopes are adjusted to achieve a sensible trade-off between estimation accuracy and computational cost. The number of partitions for the pDTDI are set to five throughout the comparison. Two benchmark problems are considered, each in two variants to assess performance under varying nonlinearity and dimensionality.

\paragraph*{Van der Pol oscillator}
The first benchmark is the discrete-time Van der Pol oscillator, representing a low-dimensional nonlinear system. This benchmark is chosen since it is well established in state estimation and has a tunable parameter $\mu$ to control the nonlinearity of the system.
Its Euler-discretized dynamics is
\[
x_{k+1}=
\begin{bmatrix}
x_{1,k}+d_Tx_{2,k}\\
x_{2,k}+d_T\big(\mu(1-x_{1,k}^2)x_{2,k}-x_{1,k}\big)
\end{bmatrix}+w_k
\]
with sampling time $d_T=0.025\,\mathrm{s}$ and nonlinearity parameter $\mu\in\{0.1,5\}$.
The measurement equation is $y_k=Cx_k+v_k$, $C=[1\;0]$.
The initial and uncertainty sets are defined as
\begin{align}
\mathcal{R}_0 = \mathcal{B}^2, \qquad
\mathcal{W} = 0.001\, \mathcal{B}^2, \qquad
\mathcal{V} = 0.2\, \mathcal{B}.
\end{align}
No control input is considered ($u_k=0$).  
Zonotope observers are evaluated with a zonotope order of 30 and PCA-based generator reduction (\cite{kopetzki_methods_2017}). Constrained zonotope methods have a maximum of 5 constraints. 

\paragraph*{Tank system}
To evaluate the scalability of the implemented observers, a nonlinear multi-tank system is considered that can be extended to an arbitrary number of interconnected tanks. 
The setup is inspired by~\cite{althoff_guaranteed_2021} and represents, for example, water levels in a cascade of hydroelectric reservoirs.
The state vector $x\in\mathbb{R}^n$ contains the water levels of all tanks, $u\in\mathbb{R}^{n_u}$ the external inflows, and $w\in\mathbb{R}^n$ the process disturbance. 
The dynamics follow Torricelli’s law, where the outflow through each orifice is proportional to the  difference of the square root of the water height between connected tanks. 
For a sampling time $d_T=0.5\,\mathrm{s}$, the Euler-discretized dynamics is
\begin{align}
x_{1,k+1} &= x_{1,k} + \tfrac{d_T}{A_1}
  \left(-\kappa_1\sqrt{2g\,x_{1,k}} + (Bu_k + w_k)_1\right),\\
x_{i,k+1} &= x_{i,k} + \tfrac{d_T}{A_i}\Big(
  \kappa_{i-1}\sqrt{2g\,x_{i-1,k}}
  - \kappa_i\sqrt{2g\,x_{i,k}} \notag\\
&\hspace{5.5em} + (Bu_k + w_k)_i\Big),
\qquad i=2,\ldots,n,
\end{align}
with gravitational acceleration $g=9.81\,\mathrm{m/s^2}$, the cross-section of the tanks $A_i = 1$ and discharge coefficients $\kappa_i=0.015$. 
The measurement model is
\[
y_k = Cx_k + v_k, \qquad v_k \in \mathcal{V},
\]
where $C$ selects the measured tank levels. The observers are tested on 6 and 30 connected tanks, the overview of the tank indices with inflow and measurement is given in Table \ref{tank_indices}.

\begin{table}[t]
\centering
\caption{Tank indices with external inflow and measurements for 6- and 30-tank systems.}
\label{tank_indices}
\renewcommand{\arraystretch}{1.05}
\setlength{\tabcolsep}{20pt}
\scriptsize
\begin{tabular}{@{}ll@{}}
\toprule
\textbf{Type} & \textbf{Tank indices}\\
\midrule
Water inflow &
1, 4, 5, 7, 9, 10, 13, 15, 16,\\
& 19, 21, 22, 25, 27, 28\\[1mm]
Water level measurement &
2, 4, 5, 7, 8, 10, 11, 13, 14,\\
& 16, 17, 19, 20, 21, 22, 23,\\
& 25, 26, 27, 28, 29\\
\bottomrule
\end{tabular}
\normalsize
\vspace{-1mm}
\end{table}

The uncertain sets are defined as
\begin{align}
\mathcal{R}_0 = 20\,\mathbf{1}_n\, \oplus 4\,\mathcal{B}^n, \quad
\mathcal{W}   = 0.001\,\mathcal{B}^n, \quad
\mathcal{V}   = 0.2\,\mathcal{B}^r
\end{align}
and the inflows $u_{j,k}$ are assumed to be known exactly.

Observers are evaluated with a zonotope order of 20 for both cases and constrained zonotope methods have a maximum of 12 and 60 constraints for the 6-dimensional and 30-dimensional case, respectively.  

\begin{remark}
  Different observer types profit from different settings. It was not possible to find a single configuration that works best for all methods. Therefore, the chosen parameters represent a compromise to achieve reasonable performance across all observers. When focusing on a specific method, further tuning of parameters can lead to improved results. Additionally, through code optimization, the computation times can be further reduced for specific methods.
\end{remark}

\subsection{Special Setups}
All proposed methods are able to handle systems of the form in (\ref{nonlinearSys}). Some, however, benefit from manual preprocessing of the nonlinear dynamics.
For DC-based observers, a programmatic way using the Hessian of the dynamics to obtain a DC split is proposed in \cite{alamo_set-membership_2008}, however, providing a manual DC decomposition strongly improves tightness.
The pDTDI observer profits from additional equality constraints~$Gx=0$ that reflect physical relationships or redundant state couplings (\cite{shen_rapid_2017}). It has been decided to apply these adaptations whenever possible without changing the behavior of the system.
For constrained zonotope observers, the improved constraint-reduction technique from \cite{rego_novel_2025} 
is used instead of the original implementation of \cite{scott_constrained_2016} to control complexity.

\paragraph*{Van der Pol oscillator}
For the VdP oscillator, the nonlinear term $\mu(1 - x_1^2)x_2$ is non-convex and therefore manually decomposed into a difference of convex (DC) functions.  
Specifically, it can be rewritten as
\[
\mu(1 - x_1^2)x_2
= \mu x_2 - \frac{\mu}{8}\!\left[(x_1^2 - 2x_2)^2 - (x_1^2 + 2x_2)^2\right].
\]
With this, the terms can be grouped into convex and concave parts to obtain the DC decomposition.
For DI-based observers, the Van der Pol system is augmented with
two additional redundant states.
The extended state thus contains both the physical variables $x = [x_1,\,x_2]^\top$ and their linear combinations $y = [y_1,\,y_2]^\top$,
resulting in a four-dimensional system.  
Equality constraints of the form $Gx = 0$ are imposed to maintain the algebraic relationships at all times,
adding two additional equality conditions.
This redundant representation provides extra coupling information to the DI observer, helping to tighten the reachable sets without altering the physical system dynamics.

\paragraph*{Tank system}
The tank system already exhibits a natural DC decomposition due to the structure of Torricelli's law. Therefore, the convex and concave terms can be identified and grouped accordingly for DC-based observers.
For DI methods, the 6-tank system is augmented with three additional states following the same idea as for the Van der Pol oscillator.
The 30-tank system is not augmented due to the already high dimensionality.

\subsection{Performance Metrics}
Two metrics quantify the performance of the estimators: computation time and estimation tightness. The average computation time per step is obtained by dividing the total runtime of the estimator by the number of simulated time steps.  
Ideally, estimation tightness would be evaluated by comparing the exact volumes of each set. However, volume computation is not feasible for all set types, especially in higher dimensions. Therefore, we compute the volume of the interval hull of the estimated reachable sets $R_k$ and then take the $n$th root and average the measure over the time steps.  
Formally, let $\mathop{hull()}$ denote the interval hull operator, then the measure is defined as
\begin{equation}
  \tilde{v} = {\frac{1}{n_\mathrm{steps}}\sum_k (vol(hull(R_k)))^{1/n}}.
\end{equation}
As an alternative measure of overall conservatism that does not favor any specific set representation, a mean-width measure~$\tilde{w}$ is introduced. Here, $N=10n$ random unit directions $d_i$ are sampled and, for each estimated set $\mathcal{R}_k$, the width of its projection onto~$d_i$ is obtained via the support function. The mean width is then computed by averaging these widths over all sampled directions and all time steps.  
Formally,
\[
\tilde{w}
= \frac{1}{N\,n_{\mathrm{steps}}}
\sum_{k=1}^{n_{\mathrm{steps}}}
\sum_{i=1}^{N}
\bigl(\rho(\mathcal{R}_k,d_i)+\rho(\mathcal{R}_k,-d_i)\bigr),
\]
where $\rho(\mathcal{R},d)=\max_{x\in\mathcal{R}} d^\top x$ denotes the 
support function of~$\mathcal{R}$ in direction~$d$.

While the mean-width measure provides a more balanced view of conservatism across different set representations, the interval volume measure is deterministic and for some applications more intuitive. Therefore, both metrics are reported.

Finally, the conservatism quantities are normalized by the best (minimum) value across observers,
\begin{equation}
  \hat{v} = \frac{\tilde{v}}{\min_j \tilde{v}_{j}}, \qquad
  \hat{w} = \frac{\tilde{w}}{\min_j \tilde{w}_j}, \qquad
\end{equation}
so that~$\hat{v} = \hat{w} = 1$ denote the best achieved performance.
Observers failing to converge are assigned~$\infty$ to preserve array consistency.

\section{Results}
\label{results}
In this section, the results of the comparison are presented and discussed. For better understanding of the involved dynamics and sets, some figures support the comparison. The main findings can be seen in the tables, which show the computation time in milliseconds, the interval volume measure $\hat{v}$ and the mean width metric $\hat{w}$ of each observer for the four investigated scenarios.

\begin{figure}[t]
    \centering
    \includegraphics[width=\linewidth]{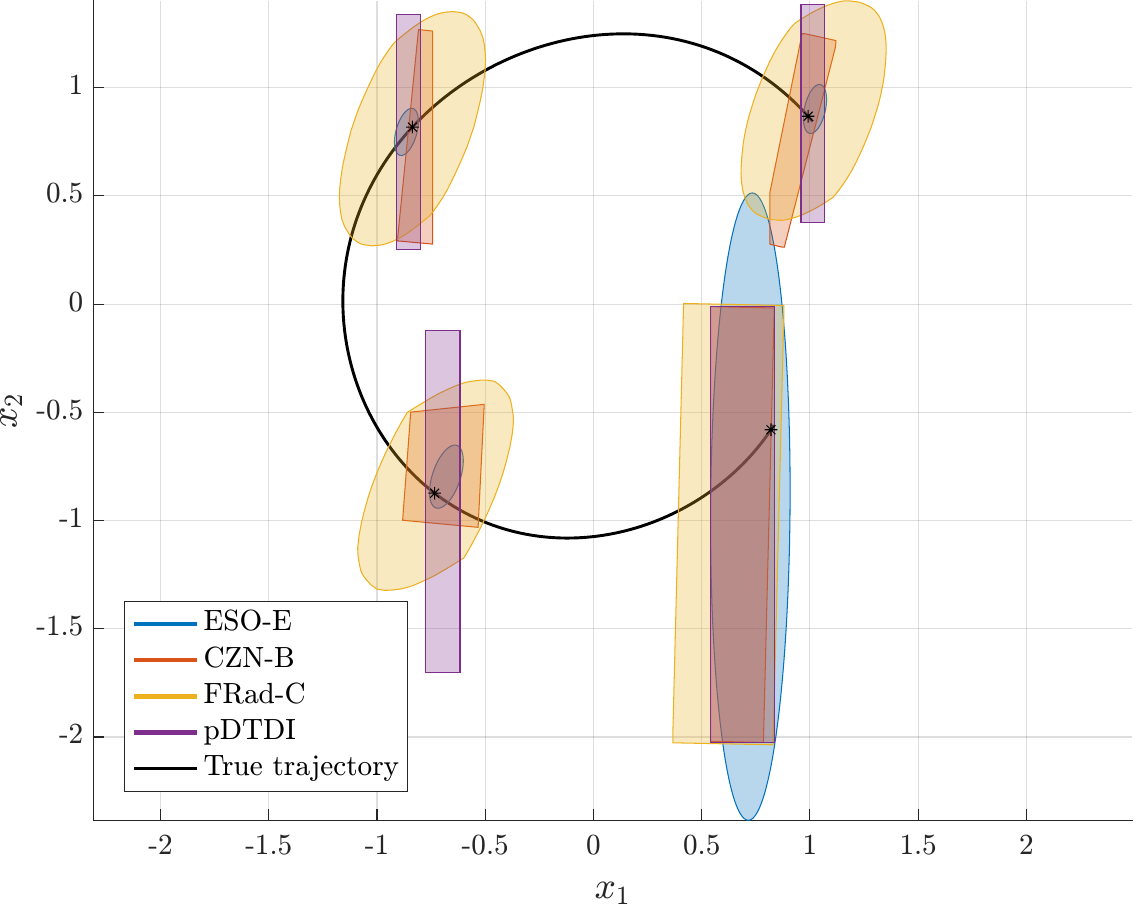}
    \caption{Reachable set enclosures in the state space for the Van der Pol oscillator with $\mu = 0.1$.}
    \label{fig:vdp_phase}
\end{figure}

Figure \ref{fig:vdp_phase} shows the dynamics of the simple Van der Pol oscillator with $\mu = 0.1$ where the generated sets at four time steps of four different observers are plotted. It can be seen that the sets decrease in size starting from the initial interval enclosures, where the ellipsoidal ESO-E method provides the tightest estimates. Interestingly, the FRad-C approach shows quite similar shapes to ESO-E, just more conservative, which is due to the high zonotope order used. For the CZN-B and pDTDI methods, it can be seen clearly which state is measured, as the sets are significantly tighter in that direction.

The full overview of the performance of all observers for this scenario is provided in Table \ref{tab:vdp_easy_results}. For this simple system, all observers are able to provide estimates for 100 iterations. The fastest method is the ZDC observer, however, all observers can run within milliseconds per step, which makes online application realistic. It can be seen clearly that the methods relying on constrained zonotopes take more time due to higher computational complexity. In terms of tightness, the ellipsoidal ESO-E method provides the best results, closely followed by the constrained zonotope methods CZDC and CZN-B. The zonotope- and interval-based methods generally show higher conservatism for this simple example.

\setlength{\tabcolsep}{14pt}
\begin{table}[ht]
  \centering
  \caption{Results for the Van der Pol oscillator with $\mu = 0.1$.}
  \label{tab:vdp_easy_results}
  \renewcommand{\arraystretch}{1} %
  \begin{tabular}{lccc}
    \toprule
    Method & Time [ms] & $\hat{v}$ & $\hat{w}$ \\
    \midrule
    \multicolumn{4}{l}{\textbf{Intersection-based Methods}}\\
    \begin{filecontents*}{tables/VdP_easy_intersection.csv}
Method,Time_ms,Interval_volume,dirWidth
ESO-E,6.55,1.00,1.00
FRad-A,3.34,2.82,2.43
FRad-B,2.28,2.80,2.46
VolMin-A,4.44,2.15,2.08
VolMin-B,3.30,3.04,4.19
ZDC,1.83,2.81,2.42
CZDC,11.27,1.55,1.46
CZN-A,19.68,2.02,1.79
CZN-B,26.90,1.12,1.21
CZMV,10.22,2.00,1.93
\end{filecontents*}
\csvreader[
      late after line=\\,
    ]{tables/VdP_easy_intersection.csv}%
    {1=\method,2=\time,3=\intvol,4=\meanWidth}%
    {\method & \MaybeInfinity{\time} & \MaybeInfinity{\intvol} & \MaybeInfinity{\meanWidth}}%
    \multicolumn{4}{l}{\textbf{Propagation-based Method}}\\
    \begin{filecontents*}{tables/VdP_easy_propagation.csv}
Method,Time_ms,Interval_volume,dirWidth
FRad-C,1.64,2.92,2.48
\end{filecontents*}
\csvreader[
      late after line=\\,
    ]{tables/VdP_easy_propagation.csv}%
    {1=\method,2=\time,3=\intvol,4=\meanWidth}%
    {\method & \MaybeInfinity{\time} & \MaybeInfinity{\intvol} & \MaybeInfinity{\meanWidth}}%
    \multicolumn{4}{l}{\textbf{Interval Methods}}\\
    \begin{filecontents*}{tables/VdP_easy_interval.csv}
Method,Time_ms,Interval_volume,dirWidth
pDTDI,11.74,1.64,2.46
ZBKH,8.22,3.37,4.66
CZKH,4.05,2.18,2.31
Best (abs.),,0.35,0.52
\end{filecontents*}
\csvreader[
      late after line=\\,
    ]{tables/VdP_easy_interval.csv}%
    {1=\method,2=\time,3=\intvol,4=\meanWidth}%
    {\method & \MaybeInfinity{\time} & \MaybeInfinity{\intvol} & \MaybeInfinity{\meanWidth}}%
    \bottomrule
  \end{tabular}
\end{table}

\begin{figure}[t]
    \centering
    \includegraphics[width=\linewidth]{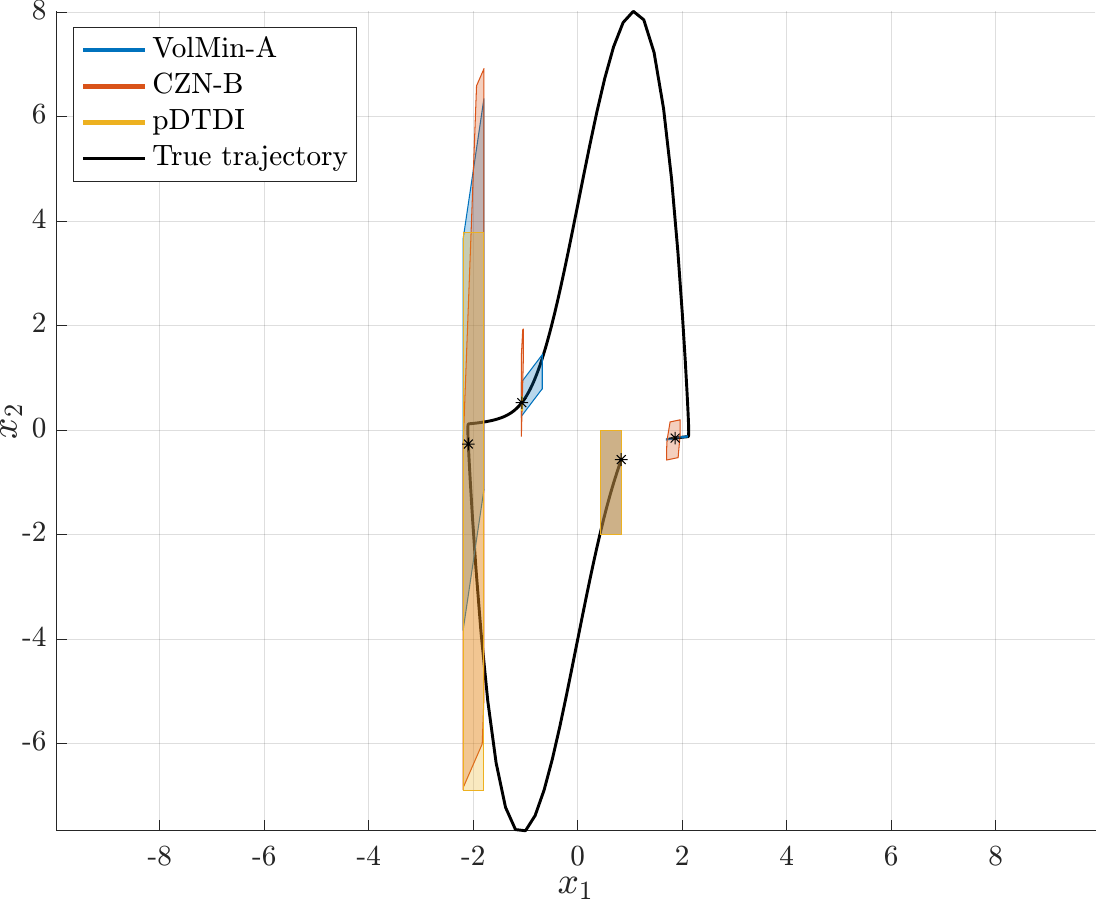}
    \caption{Reachable set enclosures in the state space for the Van der Pol oscillator with $\mu = 5$.}
    \label{fig:vdp_phase_hard}
\end{figure}

Figure \ref{fig:vdp_phase_hard} shows the effect of the increased nonlinearity on the Van der Pol oscillator ($\mu = 5$). Due to the stiff nature of the dynamics, the sizes of the estimated sets vary significantly over time. All three observers show large sets at regions with fast dynamics, especially in the not measured state variable, however, they are capable of reducing the size again quickly in slower regions.

The full results for the Van der Pol with higher nonlinearity are presented in Table \ref{tab:vdp_hard_results}. Here, some methods fail to provide estimates for the entire time horizon, indicated by infinity values. From these results we observe that the simple ellipsoidal method is not well suited for highly nonlinear systems, just like DC-based and the propagation-based method. Now the intersection-based methods using zonotopes show the fastest computation time, however the CZKH method also needs to be highlighted in that regard, showing that CZ methods can be efficient as well. Regarding the tightness, the absolute values at the bottom show, that the estimate volumes are significantly higher across all observers, compared to the simpler case. The lowest mean width is achieved by the CZN-B method here, however several methods show similar performance, like the FRad-B, VolMin-A or VolMin-B. The CZMV and ZBKH methods do not perform well in this scenario.

\begin{table}[ht]
  \centering
  \caption{Results for the Van der Pol oscillator with $\mu = 5$.}
  \label{tab:vdp_hard_results}
  \renewcommand{\arraystretch}{1}
  \begin{tabular}{lccc}
    \toprule
    Method & Time [ms] & $\hat{v}$ & $\hat{w}$ \\
    \midrule
    \multicolumn{4}{l}{\textbf{Intersection-based Methods}}\\
    \begin{filecontents*}{tables/VdP_hard_intersection.csv}
Method,Time_ms,Interval_volume,dirWidth
ESO-E,NaN,Inf,Inf
FRad-A,2.65,1.84,1.44
FRad-B,2.27,1.55,1.18
VolMin-A,4.67,1.08,1.10
VolMin-B,2.95,1.05,1.14
ZDC,NaN,Inf,Inf
CZDC,NaN,Inf,Inf
CZN-A,15.23,1.54,1.48
CZN-B,24.96,1.02,1.00
CZMV,9.60,4.02,8.00
\end{filecontents*}
\csvreader[
      late after line=\\,
    ]{tables/VdP_hard_intersection.csv}%
    {1=\method,2=\time,3=\intvol,4=\meanWidth}%
    {\method & \MaybeInfinity{\time} & \MaybeInfinity{\intvol} & \MaybeInfinity{\meanWidth}}%
    \multicolumn{4}{l}{\textbf{Propagation-based Method}}\\
    \begin{filecontents*}{tables/VdP_hard_propagation.csv}
Method,Time_ms,Interval_volume,dirWidth
FRad-C,NaN,Inf,Inf
\end{filecontents*}
\csvreader[
      late after line=\\,
    ]{tables/VdP_hard_propagation.csv}%
    {1=\method,2=\time,3=\intvol,4=\meanWidth}%
    {\method & \MaybeInfinity{\time} & \MaybeInfinity{\intvol} & \MaybeInfinity{\meanWidth}}%
    \multicolumn{4}{l}{\textbf{Interval Methods}}\\
    \begin{filecontents*}{tables/VdP_hard_interval.csv}
Method,Time_ms,Interval_volume,dirWidth
pDTDI,12.15,1.00,1.28
ZBKH,32.73,4.01,6.85
CZKH,4.34,2.21,2.63
Best (abs.),,0.95,3.21
\end{filecontents*}
\csvreader[
      late after line=\\,
    ]{tables/VdP_hard_interval.csv}%
    {1=\method,2=\time,3=\intvol,4=\meanWidth}%
    {\method & \MaybeInfinity{\time} & \MaybeInfinity{\intvol} & \MaybeInfinity{\meanWidth}}%
    \bottomrule
  \end{tabular}
\end{table}

\begin{figure}[t]
    \centering
    \includegraphics[width=\linewidth]{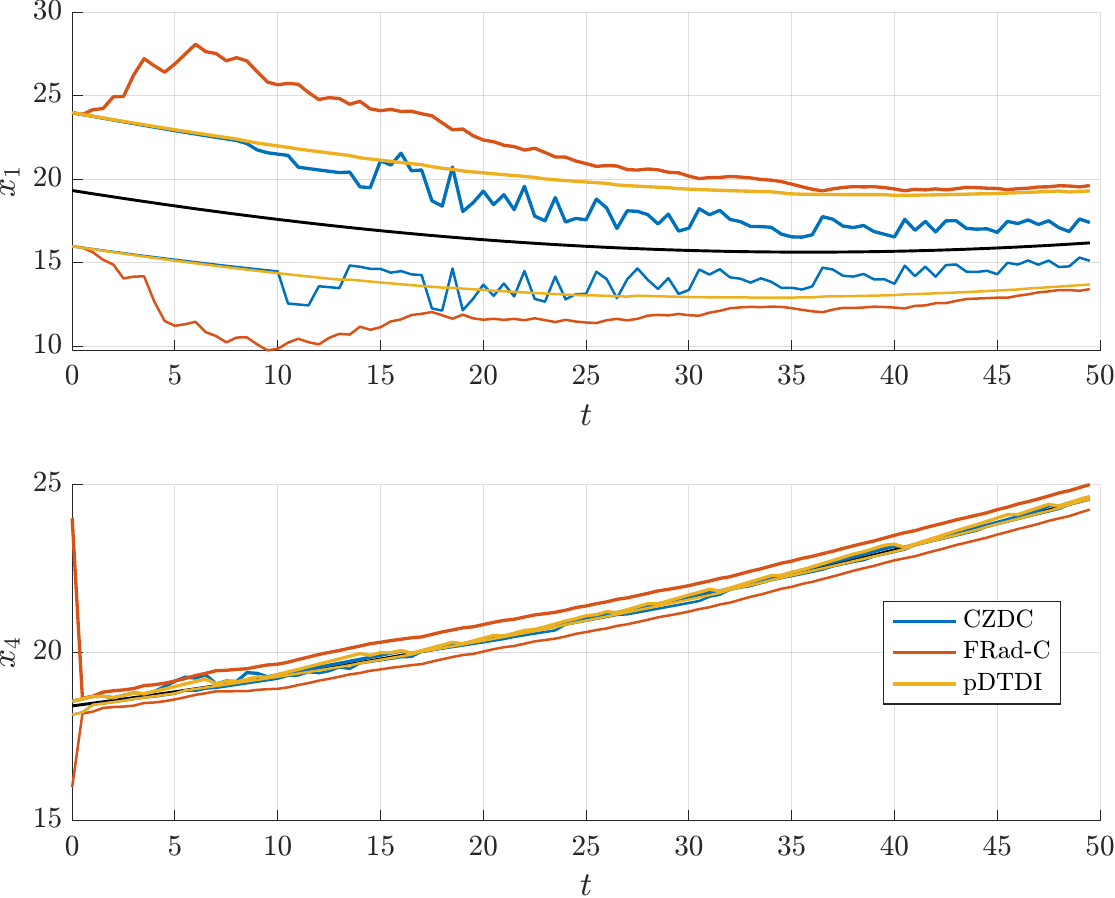}
    \caption{Interval enclosures of state variables one and four for the 6-tank benchmark system of some selected observers.}
    \label{fig:tank6}
\end{figure}

Moving on to the higher-dimensional tank systems, Figure \ref{fig:tank6} shows the interval enclosures of the state variables one and four over time for the 6-tank system of one observer of each category. The difference between measured and unmeasured state variables is clearly visible again, as the fourth state is measured and therefore shows significantly tighter estimates. CZDC provides the tightest estimates here, however, for some applications, the fluctuations in the estimates of CZDC might be undesirable, while the slightly more conservative but smoother estimates of pDTDI could be preferred.

Table \ref{tab:tank6_results} shows that most methods are able to handle a 6-dimensional system rather efficiently. Only the VolMin-A and CZN-A method fail to provide estimates for the entire time horizon for the 6-tank system. The fastest method is again the ZDC observer, however, there are several methods that can run within a few milliseconds per time step. Among the constrained zonotope methods, CZKH offers the lowest computational costs again and it can be seen that due to its splitting heuristic, the pDTDI method takes a bit longer in higher dimensions. For this problem setup, the tightest estimates are provided by the CZDC method, closely followed by the other viable constrained zonotope intersection-based methods and VolMin-B. Overall, interval-based methods show improved performance on this medium-dimensional benchmark compared to the Van der Pol example, where they tend to produce more conservative estimates in the comparison.

\begin{table}[t]
  \centering
  \caption{Results for the 6-tank benchmark system.}
  \label{tab:tank6_results}
  \renewcommand{\arraystretch}{1}
 \begin{tabular}{lccc}
    \toprule
    Method & Time [ms] & $\hat{v}$ & $\hat{w}$ \\
    \midrule
    \multicolumn{4}{l}{\textbf{Intersection-based Methods}}\\
    \begin{filecontents*}{tables/Tank6_intersection.csv}
Method,Time_ms,IntervalVolume,Rel_Width
ESO-E,8.58,4.12,2.11
FRad-A,4.59,2.96,1.85
FRad-B,3.98,2.95,1.87
VolMin-A,NaN,Inf,Inf
VolMin-B,11.60,2.12,1.31
ZDC,2.28,2.70,1.64
CZDC,29.06,1.28,1.00
CZN-A,NaN,Inf,Inf
CZN-B,94.69,1.88,1.37
CZMV,29.21,1.79,1.30
\end{filecontents*}
\csvreader[
      late after line=\\,
    ]{tables/Tank6_intersection.csv}%
    {1=\method,2=\time,3=\intvol,4=\meanWidth}%
    {\method & \MaybeInfinity{\time} & \MaybeInfinity{\intvol} & \MaybeInfinity{\meanWidth}}%
    \multicolumn{4}{l}{\textbf{Propagation-based Method}}\\
    \begin{filecontents*}{tables/Tank6_propagation.csv}
Method,Time_ms,IntervalVolume,Rel_Width
FRad-C,3.02,2.93,1.80
\end{filecontents*}
\csvreader[
      late after line=\\,
    ]{tables/Tank6_propagation.csv}%
    {1=\method,2=\time,3=\intvol,4=\meanWidth}%
    {\method & \MaybeInfinity{\time} & \MaybeInfinity{\intvol} & \MaybeInfinity{\meanWidth}}%
    \multicolumn{4}{l}{\textbf{Interval Methods}}\\
    \begin{filecontents*}{tables/Tank6_interval.csv}
Method,Time_ms,IntervalVolume,Rel_Width
pDTDI,56.87,1.00,1.36
ZBKH,17.15,2.05,1.41
CZKH,16.07,1.95,2.13
Best (abs.),,0.80,5.15
\end{filecontents*}
\csvreader[
      late after line=\\,
    ]{tables/Tank6_interval.csv}%
    {1=\method,2=\time,3=\intvol,4=\meanWidth}%
    {\method & \MaybeInfinity{\time} & \MaybeInfinity{\intvol} & \MaybeInfinity{\meanWidth}}%
    \bottomrule
  \end{tabular}
\end{table}

Finally, the results for the 30-dimensional system are summarized in Table~\ref{tab:tank30_results}. In contrast to the previous benchmarks, results are reported after 40 iterations instead of 100, since only three methods are able to provide estimates over the full horizon. These full-horizon results are shown separately in Table~\ref{tab:tank30_full_results}. The DC-based methods already fail at the first iteration, as expected due to their vertex enumeration step with exponential complexity $\mathcal{O}(2^n)$. Similarly, VolMin-A fails because the associated optimization problem becomes prohibitively large. The remaining failures are caused by excessive growth of the estimated sets over time, which leads either the state enclosure or the Jacobian intervals to leave the positive domain. As a consequence, the simulation terminates due to the underlying square-root dynamics.

\begin{figure}[t]
    \centering
    \includegraphics[width=\linewidth]{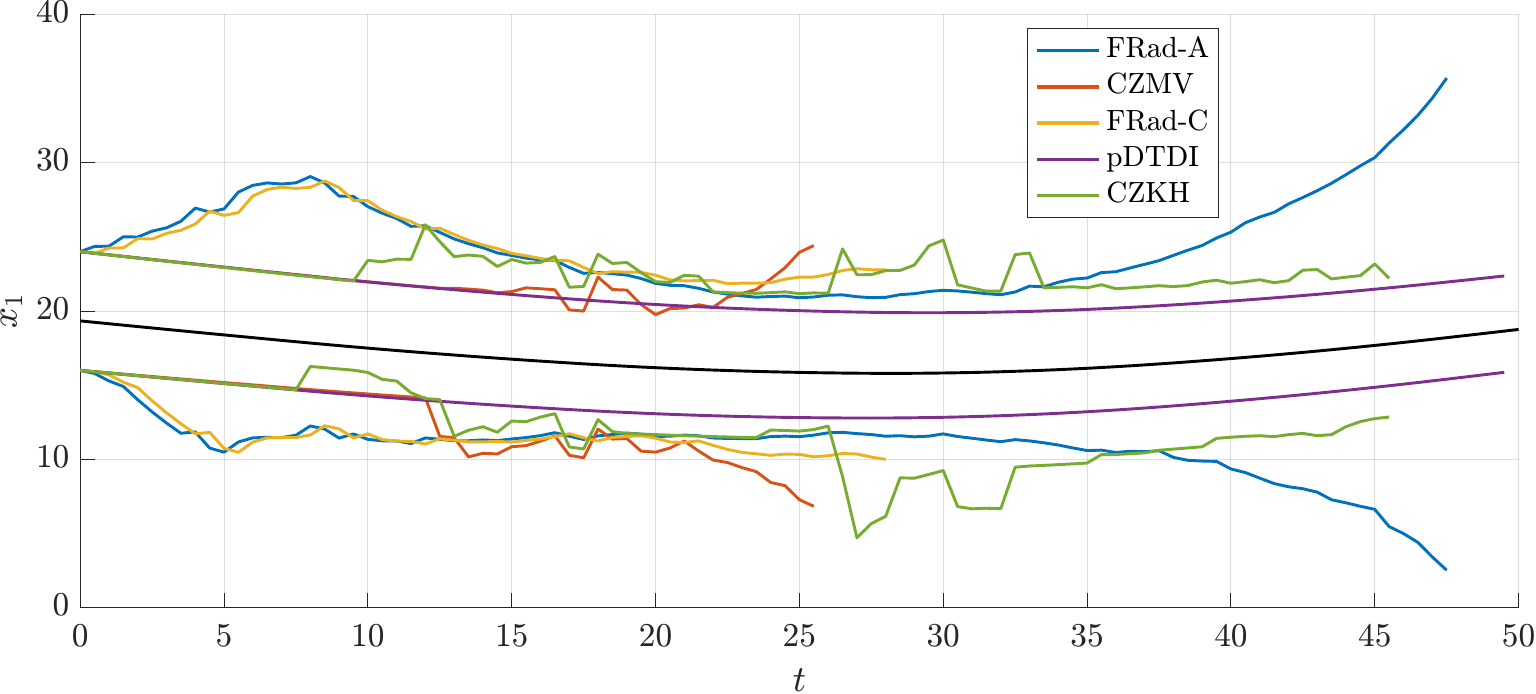}
    \caption{Interval enclosures of state variable one for the 30-tank benchmark system of some selected observers.}
    \label{fig:tank30}
\end{figure}

This behavior is illustrated in Figure~\ref{fig:tank30}, which shows the interval enclosures of the first state for several diverging observers, together with the well-performing pDTDI method as a reference. For FRad-A and CZMV, the gradual growth of the estimates is clearly visible prior to termination, whereas FRad-C and CZKH fail more abruptly.

In terms of computation time, Table~\ref{tab:tank30_results} shows a significant increase with system dimension, particularly for constrained zonotope methods. Only the approaches based on Frobenius-norm minimization maintain low computation times; however, they fail to provide tight estimates over the full horizon, which is a critical requirement for observers intended to run over long durations. The tightest enclosures are obtained by the interval-based approaches pDTDI and ZBKH, which are also among the few methods capable of providing estimates for the full horizon.

\begin{table}[t]
  \centering
  \caption{Results for the 30-tank benchmark system after 40 iterations.}
  \label{tab:tank30_results}
  \renewcommand{\arraystretch}{1}
 \begin{tabular}{lccc}
    \toprule
    Method & Time [ms] & $\hat{v}$ & $\hat{w}$ \\
    \midrule
    \multicolumn{4}{l}{\textbf{Intersection-based Methods}}\\
    \begin{filecontents*}{tables/Tank30_intersection.csv}
Method,Time_ms,IntervalVolume,Rel_Width
ESO-E,NaN,Inf,Inf
FRad-A,24.09,3.83,1.69
FRad-B,13.25,4.28,1.93
VolMin-A,NaN,Inf,Inf
VolMin-B,1126.25,3.31,1.15
ZDC,NaN,Inf,Inf
CZDC,NaN,Inf,Inf
CZN-A,NaN,Inf,Inf
CZN-B,NaN,Inf,Inf
CZMV,941.74,4.20,1.29
\end{filecontents*}
\csvreader[
      late after line=\\,
    ]{tables/Tank30_intersection.csv}%
    {1=\method,2=\time,3=\intvol,4=\meanWidth}%
    {\method & \MaybeInfinity{\time} & \MaybeInfinity{\intvol} & \MaybeInfinity{\meanWidth}}%
    \multicolumn{4}{l}{\textbf{Propagation-based Method}}\\
    \begin{filecontents*}{tables/Tank30_propagation.csv}
Method,Time_ms,IntervalVolume,Rel_Width
FRad-C,11.81,4.79,1.87
\end{filecontents*}
\csvreader[
      late after line=\\,
    ]{tables/Tank30_propagation.csv}%
    {1=\method,2=\time,3=\intvol,4=\meanWidth}%
    {\method & \MaybeInfinity{\time} & \MaybeInfinity{\intvol} & \MaybeInfinity{\meanWidth}}%
    \multicolumn{4}{l}{\textbf{Interval Methods}}\\
    \begin{filecontents*}{tables/Tank30_interval.csv}
Method,Time_ms,IntervalVolume,Rel_Width
pDTDI,367.30,1.00,1.00
ZBKH,108.36,2.71,1.06
CZKH,745.15,2.38,1.17
Best (abs.),,0.35,10.52
\end{filecontents*}
\csvreader[
      late after line=\\,
    ]{tables/Tank30_interval.csv}%
    {1=\method,2=\time,3=\intvol,4=\meanWidth}%
    {\method & \MaybeInfinity{\time} & \MaybeInfinity{\intvol} & \MaybeInfinity{\meanWidth}}%
    \bottomrule
  \end{tabular}
\end{table}

Examining the full-horizon results in Table~\ref{tab:tank30_full_results}, computation times are seen to vary significantly. While the interval methods remain within the sampling time, VolMin-B requires more than one second per time step. In terms of conservatism, pDTDI again provides the tightest estimates, followed closely by ZBKH and VolMin-B. Overall, these results highlight the difficulty of obtaining guaranteed state estimates for high-dimensional nonlinear systems.

\begin{table}[t]
  \centering
  \caption{Results for the 30-tank benchmark system after 100 iterations.}
  \label{tab:tank30_full_results}
  \renewcommand{\arraystretch}{1}
 \begin{tabular}{lccc}
    \toprule
    Method & Time [ms] & $\hat{v}$ & $\hat{w}$ \\
    \midrule
    \multicolumn{4}{l}{\textbf{Intersection-based Methods}}\\
    \begin{filecontents*}{tables/Tank30_fullintersection.csv}
Method,Time_ms,IntervalVolume,Rel_Width
VolMin-B,1256.02,4.39,1.08
\end{filecontents*}
\csvreader[
      late after line=\\,
    ]{tables/Tank30_fullintersection.csv}%
    {1=\method,2=\time,3=\intvol,4=\meanWidth}%
    {\method & \MaybeInfinity{\time} & \MaybeInfinity{\intvol} & \MaybeInfinity{\meanWidth}}%
    \multicolumn{4}{l}{\textbf{Interval Methods}}\\
    \begin{filecontents*}{tables/Tank30_fullinterval.csv}
Method,Time_ms,IntervalVolume,Rel_Width
pDTDI,326.12,1.00,1.00
ZBKH,195.28,2.94,1.04
Best (abs.),,0.31,9.97
\end{filecontents*}
\csvreader[
      late after line=\\,
    ]{tables/Tank30_fullinterval.csv}%
    {1=\method,2=\time,3=\intvol,4=\meanWidth}%
    {\method & \MaybeInfinity{\time} & \MaybeInfinity{\intvol} & \MaybeInfinity{\meanWidth}}%
    \bottomrule
  \end{tabular}
\end{table}

\section{Conclusions}
\label{conclusion}
We have compared a broad set of guaranteed state estimators for nonlinear discrete-time systems within a unified CORA framework on common benchmarks. While all methods are able to provide estimates for the Van der Pol oscillator with low nonlinearity, differences in performance become apparent as nonlinearity and dimensionality increase. DC-based methods provide tight estimates with short computation times for low- to medium-dimensional systems with mild nonlinearity, but do not scale well. The ellipsoidal method is efficient for simple systems but struggles as complexity increases. Approaches based on Frobenius-norm minimization are computationally efficient but often conservative. Volume-minimizing methods yield tight estimates in low dimensions; however, VolMin-A fails to produce results for medium- to high-dimensional systems, while VolMin-B becomes computationally expensive. Constrained zonotope methods can balance accuracy and efficiency in low to medium dimensions and can also handle strong nonlinearities, but their performance depends strongly on the system and they are not well suited for high-dimensional settings. The propagation-based method is very efficient, but it struggles with strong nonlinearities and does not yield the tightest estimates. Interval-based methods demonstrate the most robust performance across all scenarios, particularly excelling in high-dimensional systems. In particular, pDTDI appears to provide a strong baseline for practitioners, especially when dealing with higher-dimensional problems.

\begin{ack}
The authors gratefully acknowledge the financial support from the research training group ConVeY funded by the German Research Foundation under grant GRK 2428. The authors also thank the authors of the implemented methods for their valuable inputs and discussions.
\end{ack}

\section*{Declaration of Generative AI and AI-assisted technologies in the writing process}
During the preparation of this work the authors used ChatGPT 5.1 in order to refine wording and improve readability. After using this tool, the authors reviewed and edited the content as needed and take full responsibility for the content of the publication.

\bibliographystyle{elsarticle-harv}
\begin{small}
  \bibliography{autosam}
\end{small}

\end{document}